\documentclass{article}

\usepackage{arxiv}

\usepackage{amsfonts}       
\usepackage{nicefrac}       
\usepackage{microtype}      

\usepackage{float}

\usepackage{graphicx}
\usepackage[sort,numbers]{natbib}
\usepackage[T1]{fontenc}
\usepackage[cp1251]{inputenc}
\usepackage{libertine}

\usepackage{setspace}
\usepackage[colorlinks=true, citecolor=blue, linkcolor=blue, urlcolor=blue]{hyperref}
\usepackage{url}
\usepackage{booktabs}
\usepackage[justification=centering]{caption}
\usepackage{multirow}
\usepackage[version=3]{mhchem}
\usepackage{chemformula}
\usepackage{lineno}
\usepackage{lscape}
\usepackage[bottom]{footmisc}
\usepackage[affil-it]{authblk}
\usepackage{lipsum}

\newcommand\blfootnote[1]{%
	\begingroup
	\renewcommand\thefootnote{}\footnote{#1}%
	\addtocounter{footnote}{-1}%
	\endgroup
}

\title{Valorization of incinerator bottom ash for the production of resource-efficient eco-friendly concrete: Performance and toxicological characterization}

\author[1]{Ramesh B. M.} 
\author[1]{Rahul Murali Vongole}
\author[1]{Yashas Nagraj} 
\author[2]{Sujay Raghavendra Naganna}
\author[2]{Sreedhara B. M.}
\author[3]{Gireesh Mailar}
\author[4]{Ramesh P. S.}
\author[5]{Zaher Mundher Yaseen}

\affil[1]{Department of Civil Engineering, SJB Institute of Technology, Kengeri, Bengaluru - 560060, India.}
\affil[2]{Department of Civil Engineering, Siddaganga Institute of Technology, Tumakuru - 572103, India.}
\affil[3]{Department of Civil Engineering, Rural Engineering College, Hulkoti - 582205, India.}
\affil[4]{Department of Civil Engineering, Nagarjuna College of Engineering and Technology, Bengaluru - 562110, India.}
\affil[5]{New Era and Development in Civil Engineering Research Group, Scientific Research Center, Al-Ayen University, Thi-Qar, Nasiriyah, 64001, Iraq.}

\begin{document}
\maketitle
\blfootnote{Corresponding Author: E-mail: \href{sujay.gopan@gmail.com}{sujay.gopan@gmail.com}}
\begin{abstract}
\small In recent times, the quantity of wastes generated from industries, hospitals, construction sites etc. is perpetually increasing from year to year. Producing concrete utilizing waste or discarded materials as a partial replacement to fine or coarse aggregates is one among the effective ways of waste utilization. The disposal of incinerator bottom ash usually produced by the incineration of inorganic constituents of the municipal solid wastes (MSW) in an eco-friendly way is one of the issues of concern, globally. By the way, the present study is related to the utilization of MSW incinerator bottom ash and recycled demolition waste aggregate as partial replacement materials for fine and coarse aggregate, respectively to produce eco-friendly concrete. This study adopted an innovative pretreatment technique for stabilizing the MSW incinerator bottom ash. Five distinct M20 grade concrete mixes were produced with different proportions of fine aggregate, MSW incinerator bottom ash, coarse aggregate, and recycled demolition waste aggregate along with cement and water. The incinerator bottom ash was replaced at 5\% and 10\% quantities with fine aggregate and the recycled demolition waste aggregate was replaced at 40\% and 60\% of the weight of the coarse aggregate. The strength and durability properties of the M20 grade concrete were analyzed. It was noticed that the strength and durability properties of the eco-friendly concrete specimens produced by incorporating 5\% - incinerator bottom ash and 40\% - recycled demolition waste aggregate were superior to that of the control mix concrete. Laboratory tank leaching tests showed that the eco-friendly concrete do not pose any significant environmental hazard. Furthermore, the microstructural analysis through scanning electron microscope (SEM) images, revealed dense aggregate paste matrix interfaces with less micro-pores and insignificant micro-cracks due to the incorporation of incinerator bottom ash as a partial replacement to the fine aggregate.
\end{abstract}

{\small }

\section{Introduction}
\label{intro}
\par Eco-friendly Concrete is a latest terminology given to the variety of concrete produced by incorporating waste/discarded and recycled materials \citep{proske2013, revilla2020self}. Globally, several researchers have manufactured eco-friendly concrete using a wide-range of waste, recycled and composite materials \citep{xuan2016,javali2017, zareei2018, naganna2020}. The benefit of eco-friendly concrete is twofold: firstly its strength and durability properties are consistent with that of the conventional concrete, even though it includes discarded and recycled materials and secondly it contributes for waste minimization and management in a eco-friendly way \citep{javali2017}. Some of the waste or discarded materials partially replaced with cement are fly ash, silica fume, ground granulated blast-furnace slag, rice husk ash etc. Similarly, the fine aggregates are partially replaced with wastes such as spent foundry sand, coal combustion residues, bagasse ash, peanut shell ash, marble dust, waste gypsum, copper tailing, zinc tailing, lime sludge, powdered ceramic waste, waste incineration ash,  saw dust, powdered oyster shells etc., and coarse aggregates are partially replaced with discarded materials such as building demolition waste, waste glass pieces, rubber/tyre latex, crushed waste tiles, reclaimed asphalt, waste plastic chops etc. \citep{tokyay2016}. 
\par The International Solid Waste Association (ISWA), ranked India in the 3$^{rd}$ position among several countries of the world in the production of waste, particularly the municipal solid waste (MSW), with about 1.1 kilo tons of waste generated per day \citep{ISWA2013}. However, when it comes to recycling, the quantity of waste that enters into the recycling loop is very minimal and most of the wastes end up either in landfills or combustion plants. The inorganic waste materials that find way into the combustion plants in turn exit as incinerator bottom ash along with the release of toxic fumes into the air \citep{assi2020zero}. Its imprudent to bury such incinerator ash in a conventional municipal landfill \citep{giro2019rapid}. The proper management or disposal of this incinerator bottom ash is always one of the challenging issue because it requires an advanced hazard mitigating landfill which is much costlier than the conventional one \citep{ray2020utilization}. Several studies in the literature report, utilizing the incinerator ash as a substitute to fine aggregate in mass concreting works is an eco-friendly option for effective management of incinerator bottom ash \citep{joseph2018,lam2010,silva2017}. 
\par The ever increasing urban expansion in cities has led to generation of construction and demolition (C\&D) waste due to the renovation and demolition of existing structures or housing units. The demolished concrete blocks, brick masonry units, timber, metals, glass, ceramics, flooring mosaics and rubble constitute a bulk of C\&D wastes \citep{akhtar2018, wang2020design}. Currently, in India, 10-12 million tons of C\&D waste is generated annually and a larger part of these squanders are finding their way into low lying areas, tank/pond bunds, illegal dumping into stream/river course and some into landfill sites \citep{Banerjee2015}. Mass awareness, best management practices and a stringent government authority to monitor and utilize huge quantities of C\&D waste is a need of the hour in India.
\par Over the time, several studies conducted on the usage of the incinerator bottom ash as a partial replacement to fine aggregate \citep{Almuhit, pavlik2011, siddique2010, ren2020utilization, wongsa2017use} and recycled coarse aggregate (RCA) from building demolition waste as partial or complete replacement to coarse aggregate \citep{ahmed2013,mailar2017}. Zhang and Zhao (2014) \cite{zhang2014} in their investigation witnessed low chemical reactivity between the MSW incinerator bottom ash and cement in addition to an increase in the setting time (both initial and final) of the cement paste containing ash in it. In a study by Al Muhit et al. (2015) \cite{Almuhit}, the concrete incorporating MSW incinerator ash didn't gain enhanced strength and durability properties yet, held analogous properties as that of control mix samples. Some researchers who witnessed harmful components in the MSW incinerator bottom ash developed pretreatment methods for eliminating the adverse effects of reactive compounds present in the ash and thereafter managed it as a partial replacement material in concrete \citep{lynn2016,saikia2015}. Zhu et al. (2016) \cite{zhu2016} conducted field emission scanning electron microscopy (FESEM) tests to examine the heterogeneity in components of MSW incinerator bottom ash mass. Intrinsic microstructural non-uniformity was noticed in cement mortar due to the ash impurities embedded in the mortar that created cavities, micro pores and nano-cracks in the matrix hindering the adhesion of fine particles to the aggregates. The MSW incinerator bottom ash exhibited higher porosity and absorption properties in a study by Lynn et al. (2016) \cite{lynn2016}. The mortar produced by considering MSW incinerator bottom ash as a fine aggregate had reduced elastic modulus, compressive and flexural strengths when compared to that of natural aggregate mixes due to organic fraction in the ash. Liu et al. (2018) \cite{liu2018} developed an alkali treatment method for the removal of metallic aluminum from MSW incinerator bottom ash which indeed enhanced the pozzolanic properties of ash. Li et al. (2018) \cite{li2018} found that with the substitution of quartz sand with MSW incinerator bottom ash by about 60\% in autoclaved aerated concrete, the compressive strength and bulk density of concrete met the requirements of Chinese national standard. Even though recycled coarse aggregate concrete is advantageous in terms of economy and reduction in the use of natural (virgin) aggregates, some of the disadvantages include increase in water absorption capacity of concrete along with reduction in compressive strength and workability of the concrete \citep{guo2018,jain2015,yehia2015}.
\par
\begin{figure}[ht]
	\centering
	\includegraphics[scale=0.50]{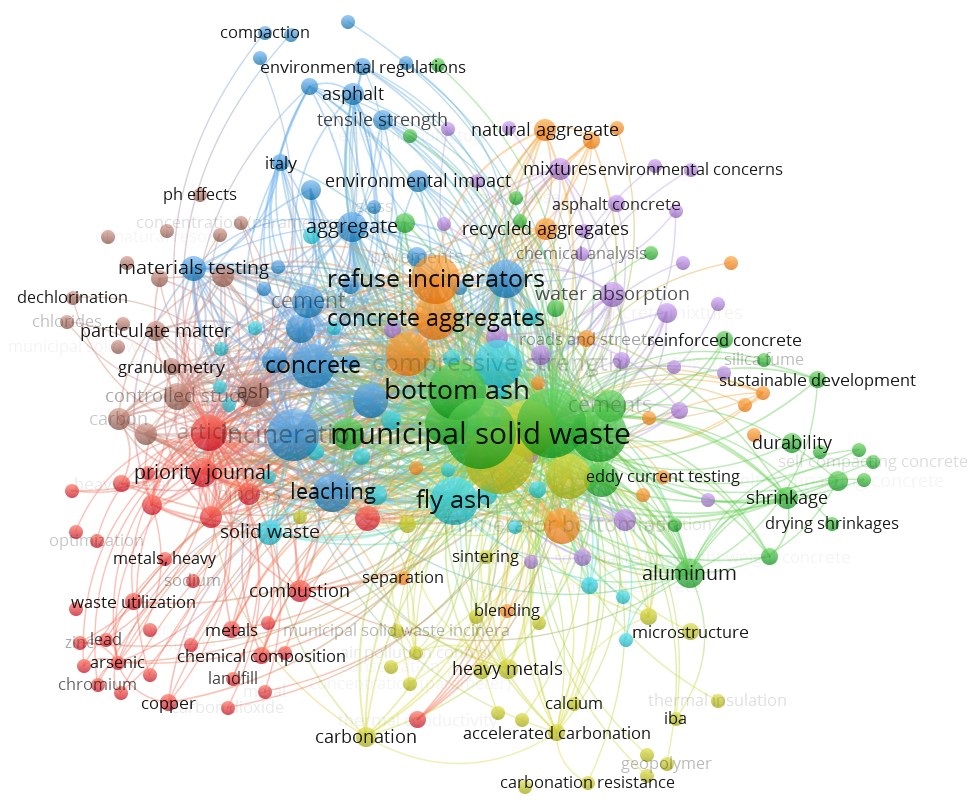}
	\caption{VOSviewer visualization of the major keywords related to literature on using MSW incinerator bottom ash as a component of concrete. The inter-relations between terms are shown as well}
	\label{fig:a}       
\end{figure}

  It is worth to visualize the reported literature on the usage of MSW incinerator bottom ash in concrete. Over 50 articles were observed in the Scopus database and the major related keywords are displayed using VOSviewer cluster algorithm as shown in Figure \ref{fig:a}. This research topic is too broad focused and different engineering aspects are explored and investigated. Among different countries, India presented minor research in this research domain and the MSW incinerator bottom ash generated in the country is a major concern for its environment. Hence, there is a provision for cautious use of MSW incinerator bottom ash as a fine aggregate in concrete. The physico-chemical properties of incinerator ash vary from one bulk to the other due to the changes or non-uniformity in the waste load shredded to the incinerator from batch to batch \citep{kuo2013use}. In addition, the MSW incinerator bottom ash is known to include several toxic residues along with heavy metals and varying amounts of moisture due to quenching \citep{wiles1996}. Further, the inconsistency in the composition of the incinerator bottom ash makes it an antagonistic material for use in concrete production \citep{clavier2019risk}. Thus, its obligatory to treat and test the physio-chemical properties of MSW incinerator bottom ash before employing it as a partial replacement material in concrete production. This is where the current research contribution and motivation are presented.
\par Based on the chemical composition of MSW incinerator bottom ash, an efficient pretreatment system has to be designed to transform it into a stable, non-reactive (inert) and homogeneous material. Hence, the objective of the present study was to design an innovative pretreatment technique for stabilizing the MSW incinerator bottom ash and thereafter utilize it as a partial replacement material to fine aggregate. The recycled demolition waste aggregate was also employed as a substitution material for coarse aggregate to produce eco-friendly concrete. This study adopted an innovative pretreatment technique for stabilizing the incinerator bottom ash. Five distinct M20 grade concrete mixes as per Indian Standard (IS) codal provisions were designed and produced with different proportions of fine aggregate (M-sand), MSW incinerator bottom ash, coarse aggregate, recycled demolition waste aggregate along with cement and water. A standard/control concrete mix and four other trial concrete mix proportions containing 5\% or 10\% of MSW incinerator bottom ash, and 40\% or 60\% of recycled demolition waste aggregate were designed. The MSW incinerator bottom ash was restricted just up to 10\% substitution taking into account of the adverse impacts that may happen such as cavities, micro-pores, micro-cracks, and leaching of heavy metals from concrete at higher replacement percentages \citep{zhu2016}. Similarly, the recycled demolition waste aggregate was replaced at 40\% and 60\% for the reason that no significant variations were observed in the strength properties of recycled aggregate concrete and the conventional concrete in the study by Robu et al. (2016) \cite{robu2016}. The strength, durability and toxicicity properties of the M20 grade concrete specimens are analyzed and reported.

\section{Materials}
\label{sec:2}

\subsection{Cement}
\label{sec:2.1}
\par Ordinary Portland Cement (OPC) conforming to IS:12269-2013 (2013) \cite{is12269:2013} of 53 grade was used throughout the experimental programme. The physical properties and chemical composition of cement as furnished by the manufacturer are presented in Table \ref{tab:1}.

\begin{table}[ht]
	\centering
	\caption{Physical and Chemical Properties of 53 Grade OPC Cement}
	\label{tab:1}

		\begin{tabular}{lll}
			\hline
			\addlinespace[0.1cm]
			\textbf{Properties} & \textbf{OPC 53 Grade} & \textbf{IS Requirement} \\ \hline
			\addlinespace[0.1cm]
			\multicolumn{3}{l}{Physical properties} \\ \hline
			\addlinespace[0.1cm]
			Specific surface ($m^{2}/kg$) & 280 & 225 min. \\
			Setting Time (minutes) &  &  \\
			Initial & 70 & 30 min. \\
			Final & 250 & 600 max. \\
			Soundness &  &  \\
			By Le-Chatelier (mm) & 3 & 10 max. \\
			By Autoclave (\%) & 0.2 & 0.8 max. \\
			Compressive strength (MPa) &  &  \\
			3 days & 35 & 27 min. \\
			7 days & 45 & 37 min. \\
			28 days & 58 & 53 min. \\ \hline
			\addlinespace[0.1cm]
			\multicolumn{3}{l}{Chemical Composition} \\ \hline
			\addlinespace[0.1cm]
			Chloride (Cl) (\%) & 0.05 max. & 0.10 max. \\
			Magnesium oxide (MgO) (\%) & 2.5 max. & 6.0 max. \\
			Sulphuric anhydride ($SO_{3}$) (\%) & 2.75 max. & \begin{tabular}[c]{@{}l@{}}2.5 max. when C3A\textless{}5;\\ 3.0 max. when C3A\textgreater{}5\end{tabular} \\
			Alumina Iron Ratio (A/F) (\%) & 1.10 min. & 0.66 min. \\
			Lime Saturation Factor (LSF) (\%) & 0.90 min. & 0.80 - 1.02 \\
			Insoluble residues (\%) & 2.0 max. & 3.0 max. \\
			Loss of Ignition (\%) & 3.0 max. & 4.0 max. \\ \hline
			\multicolumn{3}{l}{Note: min. -- minimum; max. -- maximum}
		\end{tabular}
	
\end{table}

\subsection{Aggregates}
\label{sec:2.2}
\par The manufactured sand (M sand) having particle size distribution conforming to the zone II grading of Table 4 of IS:383-1970 (2002) \cite{is383} was used as fine aggregate in the present experimental programme. The shape of M sand grains ranged from spherical to sub-angular with uniform texture. Usually the shaping machine in crushers deliver rounded/spherical shaped M sand grains free from silt, clay, and elongated or flaky particles. Some physical properties of M sand are tabulated in Table \ref{tab:2}. The pre-treated MSW incinerator bottom ash sieved through 4.75 mm IS sieve was employed as a partial replacement material for M sand. On the whole, the incinerator bottom ash is a chemically reactive material which need to be stabilized before its use. The stabilization process of MSW incinerator bottom ash is explained in detail in the next section. Meanwhile, the physico-chemical properties of stabilized MSW incinerator bottom ash used in the experimental programme are listed in Table \ref{tab:3}.  
\par Crushed granite gravel of 20 mm downsize were used in addition to recycled demolition waste (RDW) aggregate\footnote{Mortar waste was discarded/eliminated and only coarse aggregates of demolished concrete were considered for recycling} as coarse aggregates in the experimental programme of this study. Demolished concrete waste was mechanically crushed to extract the coarse aggregates from it. Visually, the coarse RDW aggregates appeared to be dense and had rough surface for strong bondage with the mortar paste. The properties of crushed granite gravel and RDW aggregates are tabulated in Table \ref{tab:2}.

\begin{table}[ht]
	\centering
	\caption{Physical Properties of Coarse and Fine Aggregates}
	\label{tab:2}
	
		\begin{tabular}{lccc}
			\hline
			\addlinespace[0.1cm]
			\multirow{2}{*}{\textbf{Property}} & \textbf{Fine Aggregate} & \multicolumn{2}{c}{\textbf{Coarse Aggregate}} \\ \addlinespace[0.1cm]\cline{2-4} 
			\addlinespace[0.1cm]
			& M sand & \multicolumn{1}{l}{Recycled Demolition Waste} & \multicolumn{1}{l}{Crushed Granite Gravel} \\ \hline
			\addlinespace[0.1cm]
			Particle size & 4.75 mm and down & 20 mm and down & 20 mm and down \\
			Fineness modulus & 2.73 & 6.19 & 6.35 \\
			Specific gravity & 2.59 & 2.70 & 2.74 \\
			Water Absorption & 0.55\% & 0.45\% & 0.33\% \\
			Bulk Density & 1510 kg/$m^3$ & 1504 kg/$m^3$ & 1545 kg/$m^3$ \\ \hline
		\end{tabular}%
	
\end{table}

\begin{table}[ht]
	\centering
	\caption{Physico-chemical properties of treated MSW Incineration bottom ash}
	\label{tab:3}
	
		\begin{tabular}{lcccc}
			\hline
			\addlinespace[0.1cm]
			\multicolumn{5}{l}{\textbf{Physical Properties}} \\ \hline \addlinespace[0.1cm]
			Property & \begin{tabular}[c]{@{}c@{}}Number of \\ samples\end{tabular} & Mean & \begin{tabular}[c]{@{}c@{}}Standard\\ Deviation\end{tabular} & Range \\ \addlinespace[0.1cm] \hline
			\addlinespace[0.1cm]
			Specific Gravity & 6 & 1.96 & 0.13 & 1.78 - 2.16 \\
			Bulk Density ($kg/m^{3}$) & 6 & 1186.67 & 126.32 & 985 - 1330 \\
			\begin{tabular}[c]{@{}l@{}}Water  Absorption (\%)  \\  $\rightarrow$  Fine Fraction  (\textless{} 4.75 mm)\end{tabular} & 6 & 12.48 & 0.48 & 11.8 - 13.20 \\
			Fineness modulus & 6 & 1.98 & 0.10 & 1.85 - 2.10 \\ \hline \addlinespace[0.1cm]
			\multicolumn{5}{l}{\textbf{Chemical Composition}} \\ \hline \addlinespace[0.1cm]
			\ce{SiO2} (\%) & 6 & 23.25 & 4.68 & 18.67 - 29.26 \\
			\ce{CaO} (\%) & 6 & 9.33 & 2.89 & 6.66 - 13.09 \\
			\ce{Fe2O3} (\%) & 6 & 15.86 & 4.85 & 10.25 - 21.12 \\
			\ce{Al2O3} (\%) & 6 & 6.65 & 2.34 & 3.76 - 9.12 \\
			\ce{Na2O} (\%) & 6 & 3.65 & 0.95 & 2.15 - 4.96 \\
			\ce{SO3} (\%) & 6 & 1.06 & 0.32 & 0.22 - 1.87 \\
			\ce{TiO2} (\%) & 6 & 0.89 & 0.18 & Nil - 1.34 \\
			\ce{K2O} (\%) & 6 & 1.56 & 0.43 & 0.72 - 2.43 \\
			\ce{MgO} (\%) & 6 & 3.28 & 0.95 & 1.68 - 4.64 \\
			\ce{P2O5} (\%) & 6 & 0.76 & 0.23 & Nil - 1.47 \\
			\ce{ZnO} (\%) & 6 & 4.55 & 1.15 & 2.06 - 5.98 \\
			\ce{KCl} (\%) & 6 & 3.76 & 1.05 & 1.27 - 5.05 \\
			\ce{PbO} (\%) & 6 & 1.68 & 0.56 & 0.45 - 2.38 \\
			\ce{MnO2} (\%) & 6 & 5.89 & 1.28 & 3.08 - 7.16 \\
			\ce{BaO} (\%) & 6 & 0.68 & 0.21 & Nil - 0.94 \\
			\ce{Cu2O} (\%) & 6 & 4.28 & 1.45 & 2.11 - 6.16 \\
			\ce{Cr2O3} (\%) & 6 & 0.75 & 0.30 & Nil - 1.07 \\
			\ce{NiO} (\%) & 6 & 2.28 & 0.76 & 0.98 - 4.55 \\
			\ce{V2O5} (\%) & 6 & 0.49 & 0.11 & Nil - 0.87 \\
			Residues (\%) & 6 & 4.68 & 2.17 & 2.03 - 6.97 \\
			LOI (\%) & 6 & 2.87 & 0.77 & 1.95 - 3.08 \\ \hline
			\multicolumn{5}{l}{Note: LOI: Loss of Ignition}
		\end{tabular}
	
\end{table}

\subsection{Pretreatment of MSW incinerator bottom ash}
\label{sec:2.3}
\par
\begin{figure}[ht]
	\centering
	\includegraphics[scale=0.65]{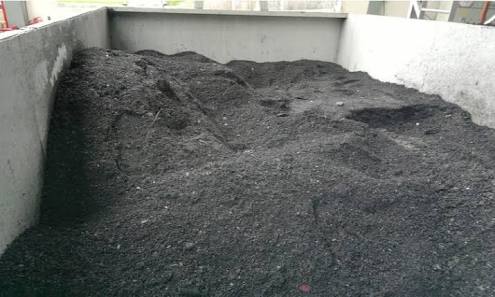}
	\caption{Bottom ash heap stored at MSW incineration plant location}
	\label{fig:1}       
\end{figure}

 The bottom ash (Figure \ref{fig:1}) expelled from MSW incineration plants owned by Gomti Incinco Pvt. Ltd., Bengaluru was used for pretreatment in the current experimental investigations. The MSW incinerator bottom ash so procured was sieved through 4.75 mm IS sieve with an objective to use it as a partial replacement material for M sand. The most economic way of preprocessing MSW incinerator bottom ash is through \textit{Wet Pretreatment}. Listed below are steps in the pretreatment procedure implemented to stabilize the MSW incinerator bottom ash.
\begin{itemize}
	\item About 3 kg of MSW incinerator bottom ash was processed at once in lab-scale batch experiments. The procedure involves sequential washing with acidic and alkaline solutions followed by cyclic water-washing process until chlorides, sulphates, soluble organic and inorganic residues are evacuated.
	\item Initially, 1M \ce{HCl} solution of about five times the quantity of ash i.e., 15 liters was taken in a PVC plastic bucket. The MSW incinerator bottom ash was slowly poured into the bucket with constant stirring of \ce{HCl} solution and thereafter the mixture was thoroughly stirred at a uniform rate for about 10 minutes. Further, the mixture was allowed to react under stable conditions for another 20 minutes. 
	\item Next, the acid-ash mixture was filtered using a nylon fine mesh filter pad and the filtrate or filter cake was sun dried by uniformly spreading over a PVC sheet.
	\item Secondly, the acid treated bottom ash was subjected to alkaline treatment by stirring and mixing it with 15 liters of 5M \ce{NaOH} solution. The mixture was again allowed to react under stable conditions for 20 minutes before filtering and later sun dried. 
	\item In the next step, the acid and alkali treated bottom ash was washed using 20 liters of deionized water by mixing and continuously stirring it in a bucket. The mixture was thoroughly stirred at a uniform rate for about 15 minutes and allowed to react under stable conditions for another 10 minutes before filtering. The washed water was tested for the concentration of chlorides and sulphates in it by titration methods.
	\item This water-washing process was carried out until the chlorides and sulphates concentration in the washed water came with in the permissible limits\footnote{The permissible limits of chloride and sulphate levels were < 250 mg/l and 200 mg/l, respectively as per IS 10500: 2012\cite{IS10500}}. Finally, the filter cake was sun dried and exposed to open environment for 24 hours and used as a partial replacement material for M sand. 
\end{itemize}

\section{Experimental Programme}
\label{sec:3}
Concrete of M20 Grade was designed as per the guidelines of IS:10262-2009 (2009) \cite{is10262} with an aim to achieve characteristic compressive strength of 20 MPa at 28 days of curing; however, the design target strength was higher by 33\% of the characteristic compressive strength. The Table \ref{tab:4} unveils the quantity of materials required to produce 1$m^3$ volume of concrete. The designed concrete mixes were expected to have a slump of 75 mm for medium degree of workability. Specimens of concrete were cast under mechanical vibration. The control mix (M1) represents the concrete produced without any  admixtures/partial replacement materials. Mixes M2 and M3 contained 40\% and 60\% replacement of crushed granite aggregate with RDW aggregate, respectively, in addition to 5\% replacement of M sand with MSW incinerator bottom ash. Similarly, mixes M4 and M5 were designed to have 10\% replacement of M sand with MSW incinerator bottom ash besides 40\% and 60\% replacement of crushed granite aggregate with RDW aggregate, respectively. Influence of MSW incinerator bottom ash and RDW aggregates on the mechanical and durability properties of concrete was studied through laboratory experiments. The concrete mixes M2, M3, M4 and M5 are hereby referred to as eco-friendly concrete.

\begin{table}[ht]
	\centering
	\caption{Mix composition of concrete mixes}
	\label{tab:4}

		\begin{tabular}{llccccc}
			\hline
			\addlinespace[0.1cm]
			\multicolumn{2}{l}{MIX} & M1 & M2 & M3 & M4 & M5 \\ \hline
			\addlinespace[0.1cm]
			Cement & Content in ($kg/m^{3}$) & 425.73 & 425.73 & 425.73 & 425.73 & 425.73 \\ \addlinespace[0.1cm]
			\multirow{2}{*}{\begin{tabular}[c]{@{}l@{}}MSW Incinerator\\ Bottom ash\end{tabular}} & Percentage & 0 & 5 & 5 & 10 & 10 \\
			& Content in ($kg/m^{3}$) & 0 & 32.25 & 32.25 & 64.49 & 64.49 \\ \addlinespace[0.1cm]
			\multirow{2}{*}{M Sand} & Percentage & 100 & 95 & 95 & 90 & 90 \\
			& Content in ($kg/m^{3}$) & 644.94 & 612.69 & 612.69 & 580.45 & 580.45 \\
			\addlinespace[0.1cm]
			\multirow{2}{*}{\begin{tabular}[c]{@{}l@{}}Crushed Granite \\ Aggregate\end{tabular}} & Percentage & 100 & 60 & 40 & 60 & 40 \\
			& Content in ($kg/m^{3}$) & 1161.73 & 697.04 & 464.69 & 697.04 & 464.69 \\
			\addlinespace[0.1cm]
			\multirow{2}{*}{\begin{tabular}[c]{@{}l@{}}Recycled Demolition\\ Waste Aggregate\end{tabular}} & Percentage & 0 & 40 & 60 & 40 & 60 \\
			& Content in ($kg/m^{3}$) & 0 & 464.69 & 697.04 & 464.69 & 697.04 \\
			\addlinespace[0.1cm]
			Water to Cement Ratio & Ratio & 0.45 & 0.45 & 0.45 & 0.45 & 0.45 \\ \hline
		\end{tabular}%
	
\end{table}

\subsection{Determination of fresh and hardened properties of eco-friendly concrete}
\label{sec:3.1}
Workability and Consistency of concrete mixes with and without partial replacement materials were examined by conducting Slump \& Vee-Bee consistometer tests as per the guidelines of IS:1199-1959 (1991) \cite{is1199}. The unconfined compressive strength test was carried out on concrete cubes of size 150 mm at the age of 3, 7, 14, 28 and 90 days at a standard loading rate as per IS:516-1959 (2006) \cite{is516}. Cylindrical specimens of 150 mm diameter and 300 mm height were cast for determining the split tensile strength of eco-friendly concrete mixes at the age of 28 days as per IS:5816-1999
220 (2004) \cite{is5816}. Concrete cast into rectangular prisms (beams) of size $100\times100\times500$ mm and cured in water for 28 days were tested for determining the flexural strength of hardened eco-friendly concrete mixes as per IS:516-1959 (2006) \cite{is516} by subjecting each specimen under a two-point loading system where the load is applied at a constant rate perpendicular to the finishing surface of the specimen. Additionally, non-destructive testing (NDT) techniques, namely the Ultrasonic Pulse Velocity (UPV) and Rebound Hammer (RH) tests were conducted as per the guidelines of IS:13311[Part-1]-1992 (2004) \cite{is133111} and IS:13311[Part-2]-1992 (2004) \cite{is13311}, respectively. The strength of concrete cube of each mix was assessed qualitatively by conducting rebound hammer tests at multiple loci of the cube (six readings of rebound number was taken from each face of cube) to arrive at average rebound number.

\subsection{Determination of durability properties of eco-friendly concrete}
\label{sec:3.2}
The durability of concrete is a direct function of its porosity and hence the Volume of Permeable Voids (VPV) test was conducted on eco-friendly concrete specimens as per the ASTM:C642-06 (2006) \cite{astm} guidelines to assess the quantum of permeable pore space in the hardened eco-friendly concrete specimens. Tank leaching tests were performed to assess the leaching of possibly hazardous substances from the monolithic eco-friendly concrete cylinders in accordance with EA NEN 7375:2004 (2005) \cite{EANEN} protocol. Trace to bulk heavy metals in the leachate samples were analyzed along with other inorganic elements (aluminium, arsenic, barium, boron, cadmium, chromium, cobalt, copper, gallium, iron, lead, manganese, nickel, silver and zinc), using high-performance inductively coupled plasma optical emission spectrometry (ICP-OES) using a \textit{Agilent Technologies} spectrometer. The leachate samples were tested for pH at time intervals of 6, 24, 54 hours and 4, 9, 19, 36 and 64 days. The permissible limits on leachable concentration of chemical contaminants as stipulated by US EPA (2010) \cite{EPA} was considered for evaluation of risk posed from leachate samples of eco-friendly concrete.

\subsection{Microstructural characterization of eco-friendly concrete}
\label{sec:3.3}
The visual examination of internal morphology and microstructure of the hydration products of eco-friendly concrete specimens was carried out by means of Scanning Electron Microscope (SEM) images captured at magnifications of 500X - 4000X.

\section{Results and Discussion}
\label{sec:4}
An attempt was made to study the interaction of MSW incinerator bottom ash with manufactured sand as fine aggregate in concrete made with blended coarse aggregates (recycled demolition waste aggregate and crushed granite gravel). The fresh state concrete properties i.e., the slump and Vee-Bee consistency of eco-friendly concrete mixes are presented in Table \ref{tab:5}. The slump values of different mixes varied between 74 mm to 90 mm indicating a "moist mix" of fair workability. Due to the addition of MSW incinerator bottom ash the viscosity and shear modulus of fresh eco-friendly concrete mixes increased facilitating a lubricant effect along with good cohesion for compactability of concrete. The pore filling effect of MSW incinerator bottom ash rendered negligible bleeding in eco-friendly concrete mixes. The consistency of concrete measured as Vee-Bee degrees (time in seconds) corresponds to the remoulding effort required to change the concrete mass from one definite shape to another. Concrete mixes with the same consistency can foster different workabilities, due to variations in the size, shape and surface texture of aggregates in the mix.

\begin{table}[ht]
	\centering
	\caption{Results of Slump and Vee-Bee Tests on Fresh Concrete}
	\label{tab:5}
	\begin{tabular}{cccc}
		\hline
		\addlinespace[0.1cm]
		MIX & Slump (mm) & Vee-Bee Degrees & Consistency \\ \hline
		\addlinespace[0.1cm]
		M1 & 74 ($\pm$ 3) & 4 & Plastic \\
		M2 & 84 ($\pm$ 3) & 3 & Semi-fluid \\
		M3 & 80 ($\pm$ 3) & 3 & Semi-fluid \\
		M4 & 90 ($\pm$ 3) & 2 & Semi-fluid \\
		M5 & 87 ($\pm$ 3) & 3 & Semi-fluid \\ \hline
	\end{tabular}%
\end{table}

\par
\begin{figure}[ht]
	\centering
	\includegraphics[scale=0.70]{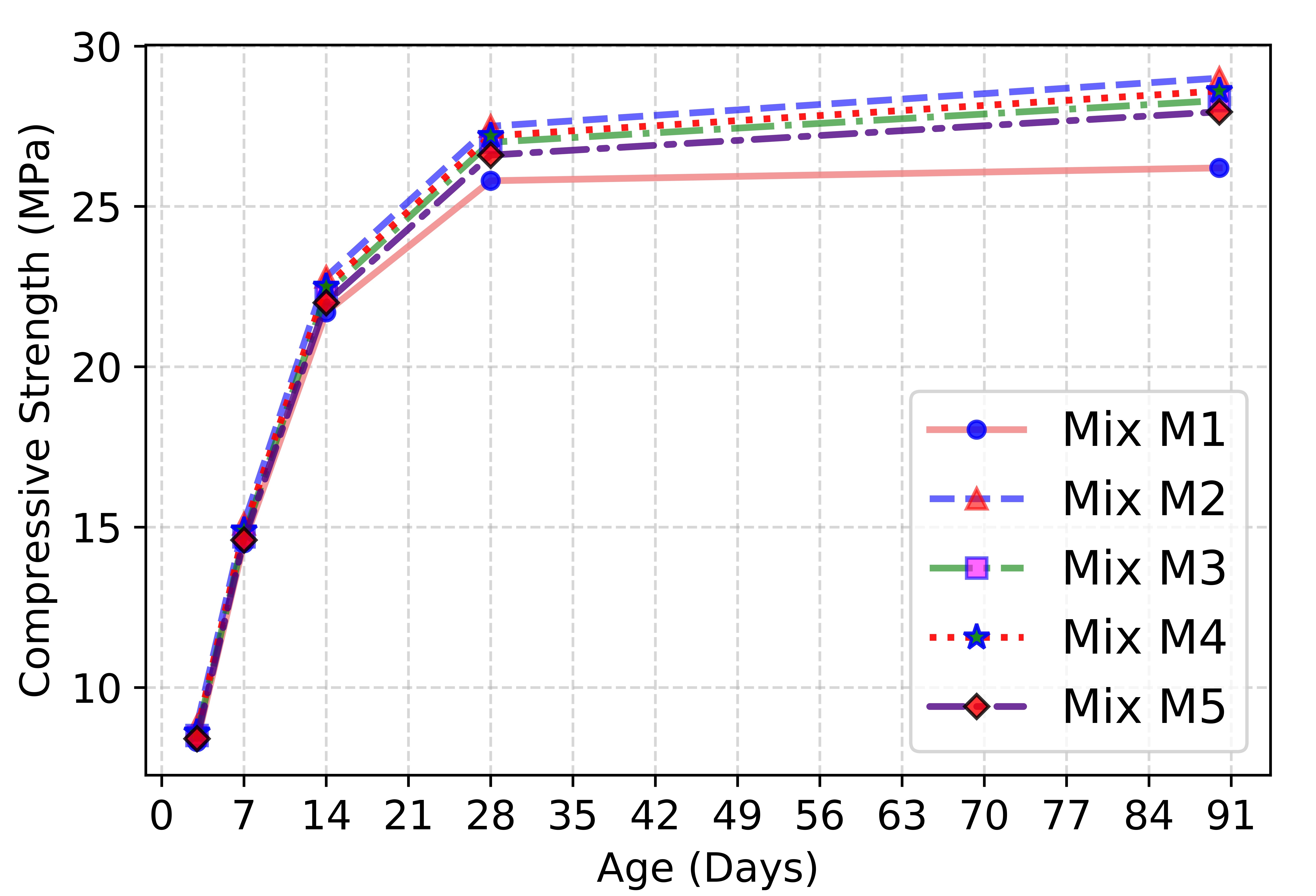}
	\caption{Compressive strength gain of the eco-friendly concrete mixes with age }
	\label{fig:2}       
\end{figure}

 Average compressive strength obtained from the testing of triplicate cubes were used to plot the strength gain curve of the eco-friendly concrete mixes with age (Figure \ref{fig:2}). The magnitude of 3 and 7 days strength of all the mixes varied relatively to each other. Significant variations in the compressive strength development of eco-friendly concrete mixes were observed at 28 and 90 days of testing.  The maximum 28 days strength of 27.5 MPa was obtained from cubes of M2 mix, this magnitude later increased to 29 MPa at 90 days of testing. In contrast, the concrete cubes of control mix (M1) had attained a strength of 25.8 MPa and 26.2 MPa at the age of 28 and 90 days, respectively. With reference to control mix (M1), the strength of eco-friendly concrete mixes M2, M3, M4, \& M5 at 90 days increased by 10.69\%, 8.02\%, 9.16\% \& 6.68\%, respectively. With an increase in MSW incinerator bottom ash content from 5\% to 10\% in the M4 and M5 concrete mixes, the strength of concrete relatively reduced owing to the interaction between any potential deleterious substances and the hydration products. In a similar manner, another finding was that with the increase in substitution of recycled demolition waste (RDW) aggregate from 40\% to 60\%, the compressive strength relatively reduced in the M3 and M5 concrete mixes. Under certain assumptions, this phenomenon could be due to intense heterogeneity within the concrete matrix and weak interfacial bonding between the cement paste and RDW aggregates. The strength gain in eco-friendly concrete mixes even after 28 days was perhaps attributed by the blending of MSW incinerator bottom ash in the mixes. It is notable that incinerator bottom ash with its chemically active silica and alumina readily reacts with the free lime to form additional insoluble and stable compounds of cementitious value.

\par The Table \ref{tab:6} below presents the split tensile and flexural strengths of eco-friendly concrete mixes at the age of 28 days. The maximum average tensile strength of 2.62 MPa was observed while testing the cylinders of M2 mix, as against 2.40 MPa strength observed in the case of control mix (M1). The tensile strength of M5 mix (containing 10\% MSW bottom ash and 60\% RDW aggregate) relatively reduced by 2.92\% of strength of control mix, owing to the degradation of paste quality by acute toxic elements of incinerator bottom ash at the interfacial transition zone. The substitution of MSW bottom ash at a dosage of 5\% in eco-friendly concrete mixes M2 and M3 demonstrated significant increase in flexural strength by 15.37\% and 10.14\% as against the reference concrete mix (M1). Likewise, the eco-friendly concrete mixes made with 40\% substitution of RDW aggregates realized an increased flexural strength performance as against the concrete mixes with 60\% RDW aggregates. Variations in the flexural strength of eco-friendly concrete mixes followed a pattern similar to that of the tensile strength behavior. On the whole, better tensile and flexural strength results achieved from M2 mix, in vivid portray it as an optimal mix combination. The optimum blend of natural crushed aggregates and recycled demolition waste aggregates increases the intact bond strength between the binder and the aggregates, and therefore increases both the tensile and flexural strengths. Additionally, when reactive aggregates like the MSW incinerator bottom ash are substituted, its free silica reacts with the excessive calcium hydroxide (one of the cement hydration product) to yield calcium silicate and other by-products which increases the overall strength of the concrete.

\begin{table}[h]
	\centering
	\caption{Split Tensile and Flexural strengths of Eco-friendly concrete mixes}
	\label{tab:6}

		\begin{tabular}{ccccc}
			\hline
			\addlinespace[0.1cm]
			MIX & \begin{tabular}[c]{@{}c@{}}Split Tensile \\ Strength (MPa)\end{tabular} & \begin{tabular}[c]{@{}c@{}}Percent\\ Increase/Decrease\end{tabular} & \begin{tabular}[c]{@{}c@{}}Flexural \\ Strength (MPa)\end{tabular} & \begin{tabular}[c]{@{}c@{}}Percent \\ Increase/Decrease\end{tabular} \\ \hline
			\addlinespace[0.1cm]
			M1 & 2.40 & - & 3.63 & - \\
			M2 & 2.62 & +9.17 & 4.19 & +15.37 \\
			M3 & 2.50 & +4.17 & 4.00 & +10.14 \\
			M4 & 2.43 & +1.25 & 3.74 & +2.98 \\
			M5 & 2.33 & -2.92 & 3.40 & -6.38 \\ \hline \addlinespace[0.1cm]
			\multicolumn{5}{l}{Note: Percent Increase/Decrease in strength is with reference to the control mix - M1}
		\end{tabular}%
	
\end{table}

\begin{table}[h]
	\centering
	\caption{Results of Non Destructive Tests  }
	\label{tab:7}
	
	\begin{tabular}{ccclcc}
		\hline
		\addlinespace[0.1cm]
		\multirow{2}{*}{MIX} & \multicolumn{2}{c}{Rebound   Hammer Test} &  & \multicolumn{2}{c}{UPV   Test} \\ \cline{2-3} \cline{5-6} 
		& \begin{tabular}[c]{@{}c@{}}Average \\ Rebound Number\end{tabular} & Strength (MPa) &  & \begin{tabular}[c]{@{}c@{}}Pulse velocity \\ (m/s)\end{tabular} & Time ($\mu$s) \\ \hline
		\addlinespace[0.1cm]
		M1 & 36 & 25.75 &  & 4020 & 36.9 \\
		M2 & 39 & 28.5 &  & 4415 & 33.9 \\
		M3 & 37 & 26.5 &  & 4320 & 34.9 \\
		M4 & 38 & 27.35 &  & 4370 & 35.4 \\
		M5 & 37 & 26.25 &  & 4235 & 36.3 \\ \hline
	\end{tabular}%
	
\end{table}

\par The test results of eco-friendly concrete cubes subjected to rebound hammer and ultrasonic pulse velocity measurements at the age of 28 days are presented in Table \ref{tab:7}. The concrete of control mix M1 reflects slightly less rebound number than the eco-friendly concrete mixes. The estimated strength values based on average rebound number were slightly equivalent to the corresponding cube compressive strength results of the eco-friendly concrete. The pulse velocity values in the range 4020--4415 m/s proclaims the quality of all eco-friendly concrete mixes as "Good" grade as per IS:13311[Part-1]-1992 (2004) \cite{is133111}. The propagation of pulse velocity through a specimen depends upon factors such as voids percentage, moisture content, density of aggregates, degree of shrinkage cracks, homogeneity and so on. There exists a strong correlation (R=0.9) between the compressive strength and pulse velocity values.

\par
\begin{figure}[ht]
	\centering
	\includegraphics[scale=0.45]{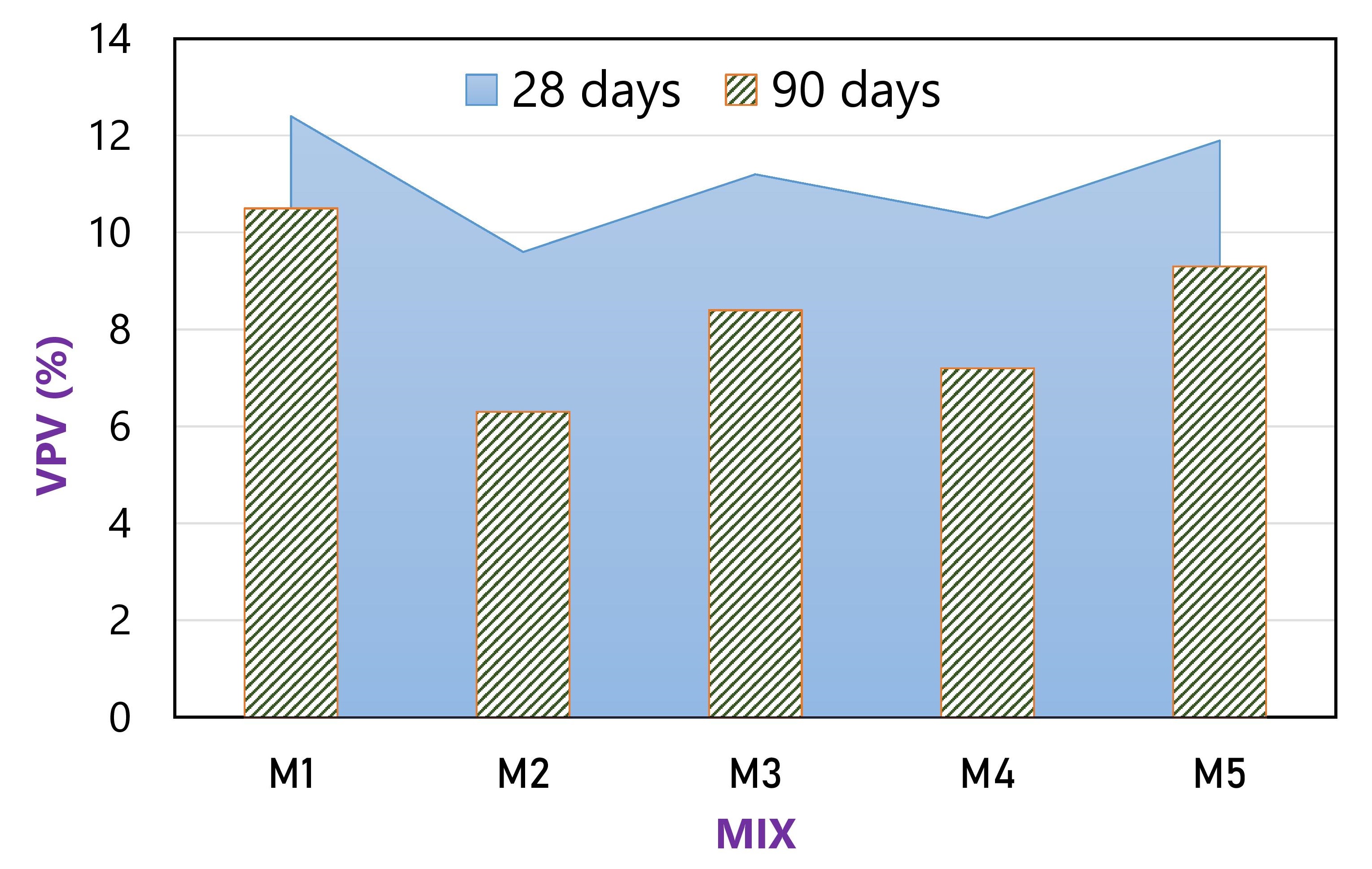}
	\caption{Volume of Permeable Voids in eco-friendly concrete specimens with age }
	\label{fig:3}       
\end{figure}

 One of the reasons for premature deterioration of concrete has been an ingress of aggressive agents and moisture through minute pores or permeable voids within concrete. Figure \ref{fig:3} shows the percent volume of permeable voids (VPV) prevailing in the eco-friendly concrete specimens assessed at the age of 28 and 90 days. It was observed that the VPV of M2 mix specimens were less by 2.8\% and 4.2\% at 28 and 90 days, respectively when compared to that of control mix M1. The eco-friendly mix M4 containing 10\% MSW bottom ash had slightly higher percentage of voids compared to M2 mix. This outcome might be due to pop-out of toxic reactions between MSW bottom ash and other hydration compounds, leading to entrapped micro-pores and interconnected voids within the concrete. The bonding defects that exist between the recycled aggregates and other concrete constituents were the reason for increase in the percentage of voids in M3 and M5 mixes. Comparably less percentage of VPV in the optimal M2 mix concrete was perhaps attributed by secondary mineralogical hydrates precipitated around the MSW bottom ash particles that fill micro-pores and lead to both pore-size and grain-size refinements.

\par
\begin{figure}[h!]
	\centering
	\includegraphics[scale=0.65]{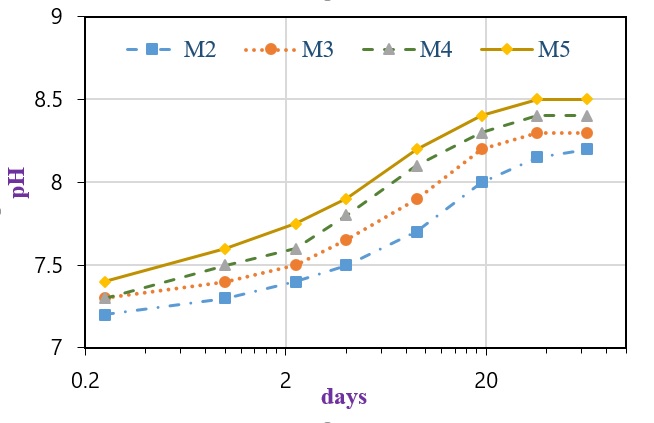}
	\caption{pH of aqueous leachate samples generated from eco-friendly concrete specimens with age}
	\label{fig:4}       
\end{figure}

 The concentration of selected elements in the aqueous leachate generated from eco-friendly concrete specimens containing MSW incinerator bottom ash was determined by  inductively coupled plasma optical emission spectrometry (ICP-OES). Table \ref{tab:8} presents the concentration of 15 toxic elements analyzed from the leachate samples of each of the eco-friendly concrete mixes (M2, M3, M4 \& M5). It is well know that under high alkaline conditions, most heavy metals get precipitated. The Figure \ref{fig:4} compares the changes in the pH of aqueous leachate samples of each eco-friendly concrete mix with age. The pH of the leachate samples increased parabolically during the initial 10 days and later followed an upward linear trend up to 64 days of leaching time. Leaching of soluble \ce{Ca(OH)2} and \ce{CaCO3} from eco-friendly concrete facilitates the growth of alkalinity in leachates samples. As Table \ref{tab:8} shows, the concentration of most of the elements analyzed were below the limits or regulated level, except the concentration of Boron which exceeds its corresponding limit in the leachate samples of M3, M4 and M5 mixes. Several catalytic reactions are known to stabilize the heavy metals contributed from incinerator bottom ash; for example, the zinc gets stabilized as calcium zincate [\ce{CaZn2(OH)5.2H2O}], and the oxyanionic chromate [\ce{CrO4^2-}] replaces \ce{SO4^2-} in ettringite (AFt) and monosulfate (AFm) products of hydration \citep{yang2018}. The elemental analysis of leachate sample of M2 eco-friendly concrete mix revealed negligible concentrations of contaminants leaching out from concrete thereby fulfilling the environmental stipulations. The pretreatment of MSW incinerator bottom ash therefore, renders an environmentally stable end-product for usage as substitution material in concrete. 

\begin{table}[h!]
	\centering
	\caption{Toxicity analysis of the expelled leachate from eco-friendly concrete specimens at the age of 64 days}
	\label{tab:8}
	\begin{tabular}{@{}ccccccc@{}}
		\toprule
		\multicolumn{1}{l}{\multirow{2}{*}{Element}} & \multirow{2}{*}{\begin{tabular}[c]{@{}c@{}}Wavelength \\ (nm)\end{tabular}} & \multicolumn{4}{c}{\begin{tabular}[c]{@{}c@{}}Concentration of toxic \\ elements (ppm)\end{tabular}} & \multirow{2}{*}{\begin{tabular}[c]{@{}c@{}}Regulated \\ level (ppm)\end{tabular}} \\ \cmidrule(lr){3-6}
		\multicolumn{1}{l}{} &  & M2 & M3 & M4 & M5 &  \\ \midrule
		\multicolumn{1}{l}{Aluminium} & 166.011 & 0.02 & 0.04 & 0.10 & 0.13 & 0.2 \\
		\multicolumn{1}{l}{Aresnic} & 188.905 & 0.32 & 0.48 & 0.66 & 0.72 & 5.0 \\
		& 189.080 & 0.04 & 0.07 & 0.11 & 0.15 & 5.0 \\
		& 201.344 & 0.01 & 0.02 & 0.06 & 0.09 & 5.0 \\
		\multicolumn{1}{l}{Barium} & 223.557 & 1.56 & 1.96 & 3.05 & 3.33 & 100 \\
		& 479.903 & 1.22 & 1.54 & 3.26 & 3.65 & 100 \\
		& 483.008 & 1.36 & 1.88 & 2.87 & 3.11 & 100 \\
		& 604.631 & 0.98 & 1.32 & 2.54 & 2.97 & 100 \\
		\multicolumn{1}{l}{Boron} & 180.457 & 1.10 & 1.65 & 2.65 & 3.66 & 1.4 \\
		& 217.909 & 0.98 & 1.30 & 2.43 & 3.28 & 1.4 \\
		& 239.680 & 0.97 & 1.21 & 2.22 & 3.03 & 1.4 \\
		& 249.632 & 0.97 & 1.18 & 2.03 & 2.95 & 1.4 \\
		\multicolumn{1}{l}{Cadmium} & 203.466 & bdl & bdl & bdl & bdl & 1.0 \\
		& 234.753 & bdl & bdl & bdl & bdl & 1.0 \\
		& 235.862 & bdl & bdl & bdl & bdl & 1.0 \\
		& 243.452 & bdl & bdl & bdl & bdl & 1.0 \\
		\multicolumn{1}{l}{Chromium} & 200.060 & bdl & bdl & 0.21 & 0.35 & 5.0 \\
		& 219.658 & bdl & bdl & 0.15 & 0.22 & 5.0 \\
		& 319.235 & bdl & bdl & 0.09 & 0.18 & 5.0 \\
		& 353.808 & bdl & bdl & 0.05 & 0.11 & 5.0 \\
		\multicolumn{1}{l}{Cobalt} & 230.305 & bdl & 0.05 & 0.25 & 0.28 & 1.0 \\
		& 203.886 & bdl & 0.03 & 0.14 & 0.20 & 1.0 \\
		& 225.433 & bdl & 0.03 & 0.06 & 0.11 & 1.0 \\
		\multicolumn{1}{l}{Copper} & 243.100 & bdl & bdl & 0.15 & 0.2 & 1.0 \\
		\multicolumn{1}{l}{- Ionic form 1} & 210.004 & bdl & bdl & 0.11 & 0.17 & 1.0 \\
		& 229.823 & bdl & bdl & 0.08 & 0.16 & 1.0 \\
		\multicolumn{1}{l}{- Ionic form 2} & 331.675 & 0.14 & 0.19 & 0.21 & 0.22 & 1.0 \\
		\multicolumn{1}{l}{Gallium} & 276.443 & bdl & bdl & bdl & bdl & 1.0 \\
		& 289.658 & bdl & bdl & bdl & bdl & 1.0 \\
		\multicolumn{1}{l}{Iron} & 236.104 & bdl & 0.04 & 0.10 & 0.12 & 0.3 \\
		& 236.883 & bdl & 0.04 & 0.09 & 0.12 & 0.3 \\
		& 251.432 & bdl & 0.02 & 0.09 & 0.07 & 0.3 \\
		& 260.670 & bdl & 0.02 & 0.06 & 0.07 & 0.3 \\
		\multicolumn{1}{l}{Lead} & 213.663 & bdl & 0.17 & 0.21 & 0.44 & 5.0 \\
		& 254.487 & 0.05 & 0.13 & 0.19 & 0.32 & 5.0 \\
		\multicolumn{1}{l}{Manganese} & 261.630 & bdl & bdl & bdl & bdl & 0.05 \\
		& 267.542 & bdl & bdl & bdl & bdl & 0.05 \\
		\multicolumn{1}{l}{Nickel} & 215.643 & bdl & bdl & 0.04 & 0.06 & 1.0 \\
		& 226.303 & bdl & bdl & 0.03 & 0.05 & 1.0 \\
		\multicolumn{1}{l}{Silver} & 241.379 & bdl & bdl & bdl & bdl & 5.0 \\
		& 252.216 & bdl & bdl & bdl & bdl & 5.0 \\
		& 333.458 & bdl & bdl & bdl & bdl & 5.0 \\
		\multicolumn{1}{l}{Zinc} & 210.220 & 0.03 & 0.07 & 0.13 & 0.24 & 1.5 \\
		& 219.457 & 0.01 & 0.06 & 0.13 & 0.21 & 1.5 \\ \midrule
		\multicolumn{7}{l}{Note: bdl -- below detection limit}
	\end{tabular}
	
\end{table}

\par
\begin{landscape}

	\begin{figure}[h!]
		\centering
		\includegraphics[scale=0.4]{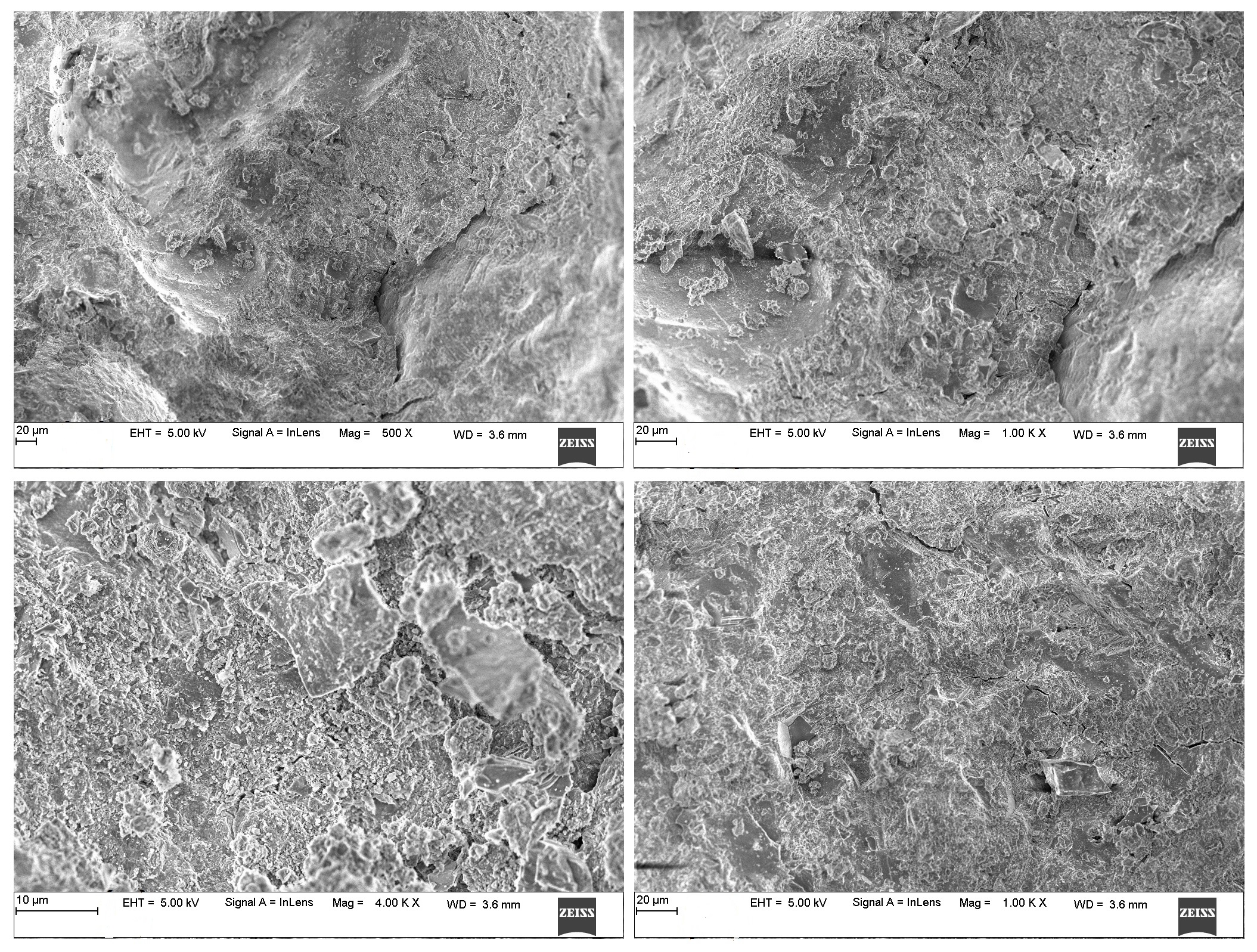}
		\caption{SEM images depicting inhomogeneity in the microstructure of eco-friendly concrete}
		\label{fig:5}       
	\end{figure}

\end{landscape}

 The inhomogeneities in the microstructure of concrete is generally accompanied by the formation of internal micro-cracks which, when subjected to tensile stresses caused by weathering and loading effects during the service life, gradually grow into surface defects \citep{proc-2005}. The SEM micrographs presented in Figure \ref{fig:5} reveals that, the partial replacement of MSW incinerator bottom ash with M sand in eco-friendly concrete mixes imparts a non-uniform compact pore structure system with multiphase products primarily of calcium silicate hydrate (C-S-H) and calcium alumino sulphate hydrates. However, micro-cracks that look insignificant, typically less than 10$\mu$m were observed at interfacial transition zone (ITZ) between paste matrix and fine aggregates. The heavy metals immobilize through chemical adsorption and multiple isomorphous replacement in hydrated silicate or aluminate phases \citep{fan2018}.

\section{Conclusions}
\label{sec:last}
The eco-friendly concrete produced by utilizing MSW incinerator bottom ash and recycled demolition waste aggregates proved to be an excellent alternative to conventional concrete. The increase in compressive strength of eco-friendly concrete even after 28 days, makes it acceptable for all structural applications. The optimal replacement percentage of MSW incinerator bottom ash and recycled demolition waste aggregates was 5\% and 40\%, respectively to attain significant strength and durability properties. The percent volume of permeable voids in specimens of eco-friendly concrete mixes were much lower compared to control mix concrete. The pretreatment of MSW incinerator bottom ash rendered an environmentally stable end-product for usage as substitution material in concrete. Laboratory tank leaching tests showed that the eco-friendly concrete do not pose any significant environmental hazard. Furthermore, the microstructural analysis revealed dense aggregate paste matrix interfaces with less micro-pores and insignificant micro-cracks due to the incorporation of incinerator bottom ash as a partial replacement to the fine aggregate. The determination of sulphates and chlorides in the eco-friendly concrete could be considered as a future scope of work.

\section*{CRediT author statement}
Ramesh B. M: Conceptualization, Methodology, Resources; Sujay Raghavendra Naganna: Writing - Original draft, Visualization; Sreedhara B. M: Writing - Original Draft; Gireesh Mailar: Validation, Data Curation; Ramesh P. S: Supervision; Rahul Murali Vongole: Investigation, Resources; Yashas Nagraj: Investigation, Resources; Zaher Mundher Yaseen: Writing - Reviewing and Editing.

\section*{Acknowledgements}
The authors wish to acknowledge the cooperation rendered by the non-teaching staff of Department of Civil Engineering, SJBIT, Bengaluru and the management of SJBIT, Bengaluru for the necessary infrastructural support for obtaining the experimental data presented in the
manuscript.

\section*{Disclosure of information and Conflict of interest}
The authors confirm that there are no known conflicts of interest associated with this publication and there has been no financial gains for this work that could have influenced its outcome.

\section*{Funding}
This research received no external funding.

\bibliographystyle{unsrt}  
\bibliography{refer}  

\end{document}